\documentclass[conference]{IEEEtran}
\IEEEoverridecommandlockouts
\usepackage{cite}
\usepackage{amsmath,amssymb,amsfonts}
\usepackage{algorithmic}
\usepackage[dvipdfmx]{graphicx}
\usepackage{textcomp}
\usepackage{xcolor}
\usepackage{bm}
\usepackage{algorithm}
\def\BibTeX{{\rm B\kern-.05em{\sc i\kern-.025em b}\kern-.08em
    T\kern-.1667em\lower.7ex\hbox{E}\kern-.125emX}}

\begin{document}

\title{Physics-Aware Decoding for Communication Channels Governed by Partial Differential Equations   
%\thanks{This work was supported by JSPS KAKENHI Grant Number JP22H00514.}
}

\author{
\IEEEauthorblockN{Tadashi Wadayama}
\IEEEauthorblockA{ 
\textit{Nagoya Institute of Technology}\\
wadayama@nitech.ac.jp}
\and
\IEEEauthorblockN{Koji Igarashi}
\IEEEauthorblockA{ 
\textit{Osaka University}\\
iga.koji.es@osaka-u.ac.jp}
\and
\IEEEauthorblockN{Takumi Takahashi}
\IEEEauthorblockA{ 
\textit{Osaka University}\\
takahashi@comm.eng.osaka-u.ac.jp}
\thanks{This work was partly supported by a research grant from NEC cooperation.}

}

\maketitle

\begin{abstract}
Digital communication systems inherently operate through physical media governed by partial differential equations (PDEs). In this paper, we introduce a physics-aware decoding framework that integrates gradient descent-based error correcting algorithms with PDE-based channel modeling using differentiable PDE solvers. At the core of our approach is gradient flow decoding, which harnesses gradient information directly from the PDE solver to guide the decoding process. We validate our method through numerical experiments on both the heat equation and the nonlinear Schrödinger equation (NLSE), demonstrating significant improvements in decoding performance. The implications of this work extend beyond decoding applications, establishing a new paradigm for physics-aware signal processing that shows promise for various signal detection and signal recovery tasks.
\end{abstract}


\section{Introduction}
 
Digital communications and storage systems form the backbone of our modern information society. These systems invariably rely on physical media governed by fundamental laws of physics. The behavior of such physical systems is mathematically described by partial differential equations (PDEs), which capture the spatial and temporal evolution of the underlying physical phenomena. For instance, wireless communication systems utilize electromagnetic waves that follow Maxwell's equations, which describe how electric and magnetic fields propagate, interact, and evolve over time and space. In optical fiber communications \cite{Agrawal}, signal propagation is governed by the nonlinear Schrödinger equation (NLSE), a PDE that describes how the optical pulse shape and its phase evolves along the fiber length due to various physical effects such as dispersion 
	and nonlinearity.
	
In communication systems, physical phenomena are typically described by wave equations, diffusion equations, or more complex coupled systems of PDEs such as Maxwell's equations. The ability to solve these equations numerically is essential for accurate channel modeling and performance optimization. This is particularly evident in {optical fiber communications}, where researchers have long developed sophisticated signal processing techniques that explicitly account for the underlying NLSE in their channel models \cite{Agrawal}. In the area of molecular communication, which aims to enable communication between nanoscale devices using molecules as information carriers, diffusion equations play a significant role in describing the propagation of molecular signals through fluid media \cite{Farsad}. The stochastic nature of molecular diffusion, combined with the underlying physics governed by PDEs, presents unique challenges in signal detection and channel estimation. 

In the realm of digital communications, error correcting coding, such as low-density parity-check (LDPC) codes \cite{Gallager63}, 
 has been a crucial technology to ensure reliable data transmission and storage. Traditional decoding algorithms, however, have been developed primarily from an information-theoretic perspective, with limited consideration of the underlying physics of the communication channels represented by these PDEs. 
%A general framework for integrating physical models into error correction decoding remains to be established.

Recent advances in scientific computing have demonstrated remarkable efficiency in solving PDEs through machine learning approaches, particularly physics informed neural networks (PINNs) \cite{Raissi} and automatic differentiation (AD) techniques \cite{Baydin2018}. These methods achieve superior performance across various physics-based applications by directly incorporating physical constraints into the learning process. Beyond solving forward problem, PINNs have proven particularly effective for PDE-related inverse problems, highlighting the significant potential of integrating physical models with machine learning methodologies.
%Error correction with underlying physics constraint can be seen as a 
%PDE-related inverse problem \cite{Raissi}.

Despite these developments such as PINNs, the integration of numerical PDE solvers \cite{Bartels} with  decoding algorithms remains largely unexplored as a general framework. This presents a promising research direction, as the physical constraints governing the communication channel could potentially provide valuable information for the decoding process. 
	A comprehensive approach that combines {differentiable PDE solvers} with  
	decoding algorithms could benefit a wider range of communication systems.

In this paper, we introduce a novel approach to {\em physics-aware decoding} that integrates error correction techniques with PDE-based channel modeling. Building upon the foundation of {\em gradient flow (GF) decoding} \cite{wadayama23, wadayama24-2}, our method explicitly incorporates the physical constraints of communication channels as described by PDEs. The key contributions of this work are as follows:
We introduce the concept of physics-aware decoding and develop a novel decoder architecture that seamlessly integrates differentiable PDE solvers with gradient-based decoding algorithms.
We also validate our framework through numerical experiments on two different PDEs, {the heat equation and the NLSE}, demonstrating enhanced decoding performance through the effective utilization of gradient information from a PDE solver.

\section{Preliminaries}

%\subsection{Related works}

\subsection{Notation}

Assume that a binary parity check matrix $\bm H = \{H_{ij}\} \in \mathbb{F}_2^{m \times n}$ 
is given.
The binary linear code $\tilde C(\bm H)$ is defined by 
\begin{align}
\tilde C(\bm H) \equiv \{\bm{b}  \in \mathbb{F}_2^n \mid  \bm H \bm{b} = \bm{0} \}.		
\end{align}
The binary to bipolar transform $\beta: \mathbb{F}_2 \rightarrow \{1, -1\}$
defined by $\beta(0) \equiv 1$ and $\beta(1) \equiv -1$ transforms 
$\tilde C(\bm H)$
into the bipolar code defined by
\begin{align}
 C(\bm H) \equiv \{\beta(\bm{b}) \in \{1, -1\}^n \mid \bm{b}   \in \tilde C(\bm H) \}.		
\end{align}
The index sets $A(i)$ and $B(j)$ are defined as
\begin{align}
	A(i) &\equiv \{j \mid j \in [n], H_{i, j} = 1 \}, i \in [m], \\
	B(j) &\equiv \{i \mid i \in [m], H_{i, j} = 1   \}, j \in [n],
\end{align}
respectively.
The notation $[a]$ represents the set of consecutive positive integers from $1$ to $a \in \mathbb{N}$.
Note that the similar notation $[a,b]$ denotes the closed real interval from $a \in \mathbb{R}$ to $b \in \mathbb{R}$. 
A function $f:\mathbb{R} \rightarrow \mathbb{R}$ can be applied to a vector $\bm x \in \mathbb{R}^n$ as
$f(\bm x) \equiv (f(x_1),f(x_2),\ldots, f(x_n))$ where $\bm x = (x_1,\ldots, x_n) \in \mathbb{R}^n$.
Namely, a scalar function $f$ can be component-wisely  applicable to a vector $\bm x$.
For a pair of vectors $\bm a = (a_1,\ldots, a_n) \in \mathbb{R}^n, \bm b \equiv (b_1,\ldots, b_n) \in \mathbb{R}^n$, we define the multiplication and division on two vectors by 
$\bm a \odot \bm b \equiv (a_1 b_1, \ldots, a_n b_n)$ and 
$\bm a / \bm b \equiv (a_1/b_1, \ldots, a_n/b_n)$, respectively.

\subsection{Gradient Flow (GF) Decoding}

The generalized GF decoding proposed in \cite{wadayama24-2} 
is defined as follows.
The ODE for the GF decoding is given by 
\begin{align}
	\frac{d\bm x}{dt} = - (\nabla L(\bm x;\bm y ) + \gamma \nabla h_{\alpha,\beta}(\bm x)), 
\end{align}
where the initial condition is $\bm x(0) = \bm x_0$ and $\bm y$ represents the received vector.
We thus classified GF decoding into a class of optimization-based decoding 
algorithms \cite{Feldman03,Vontobel08,Zhang13,Barman13}.
The function $L(\bm x;\bm y )$ represents the negative log likelihood of the channel probability 
density function $p(\bm y | \bm x)$.
The {\em code potential energy function} for $C(\bm H)$ is a multivariate polynomial defined as
\begin{equation} \label{heq}
	h_{\alpha,\beta}(\bm{x}) \equiv \alpha \sum_{j = 1}^n (x_j^2 - 1)^2 
	+ \beta \sum_{i = 1}^m \left( \left(\prod_{j \in A(i)} x_j \right)  - 1 \right)^2, 
\end{equation}
where $\bm x = (x_1, \ldots, x_n)^T\in \mathbb{R}^n$.
The parameters $\alpha \in \mathbb{R}_+$ and $\beta \in \mathbb{R}_+$ control the 
relative strength of the first and second terms.
The first term on the right-hand side of (\ref{heq}) 
represents the bipolar constraint for $\bm{x} \in \{+1, -1\}^n$,
and the second term corresponds to the parity constraint induced by $\bm H$, i.e.,
if $\bm x \in C(\bm H)$, we have
$\left(\prod_{j \in A(i)} x_j \right) -1 = 0$
 for any $i \in [m]$.

The code potential energy $h_{\alpha,\beta}(\bm{x})$ is inspired by 
the non-convex objective function introduced in \cite{wadayama10}.
The sum-of-squares form of (\ref{heq}) directly implies the most important property 
of $h_{\alpha,\beta}(\bm{x})$, i.e., 
the inequality 
$
		h_{\alpha,\beta}(\bm{x})  \ge 0
$
holds for any $\bm{x} \in \mathbb{R}^n$. The equality holds if and only if $\bm{x} \in C(\bm H)$.

The gradient of the potential energy \cite{wadayama24-2} is required for a GF decoding process.
The gradient is given by
\begin{align}
\nabla h_{\alpha,\beta}(\bm{x}) =4 \alpha (\bm{x}\odot\bm{x}-\boldsymbol 1)\odot\bm{x}+{2\beta\boldsymbol{H}^{T}(\boldsymbol d\odot\boldsymbol d-\boldsymbol d)}/{\bm{x}},
\end{align}
where $\bm d$ is defined by
\begin{align}
\boldsymbol d \equiv (\boldsymbol 1-2\mathrm{bmod}(\boldsymbol {H}(\boldsymbol 1-\mathrm{sgn}(\bm{x})/2)))\odot\exp(\boldsymbol H\ln(|\bm{x}|)).	
\end{align}
The function $\mbox{sgn}$ is the sign function and $\mathrm{bmod}$ represents the real remainder modulo $2$ function, defined as 
\begin{align}
    \mathrm{bmod}(a)\equiv a-2\left\lfloor \frac{a}{2}\right\rfloor,\ a\in\mathbb R.	
\end{align}
%This expression of the gradient is useful for 
%deep unfolding \cite{LISTA,9020494}.

%Discretizing the time axis, we can obtain 
%the  recursive formula of the discretized GF decoding
% given by 
%\begin{align} 
%	\bm s^{(k+1)} &= \bm s^{(k)} - \eta (\nabla L(\bm s^{(k)};\bm y ) + \gamma \nabla h_{\alpha,\beta}(\bm s^{(k)})),\ k = 0,1,\ldots,
%\end{align}
%where $\bm s^{(0)} = \bm x_0$. 
%This recursive process can be regarded as a 
%gradient descent method.

\subsection{Finite Difference Methods (FDM)}

Among the various numerical methods for solving PDEs, finite difference methods (FDM) \cite{Bartels} represent the simplest approach. For clarity of presentation, we focus on the heat PDE:
\begin{align} \label{difusion_PDE}
\frac{\partial u}{\partial t}  = \lambda \frac{\partial^2 u}{\partial x^2},\quad t \in [0,T],\quad x \in [0,L].
\end{align}
By applying forward differences to the temporal derivative and central differences to the spatial derivative, we obtain the following difference equation:
\begin{align}
u(t+h, x) = (1 - 2c) u(t,x) + c u(t, x + \ell) + cu(t, x - \ell),	
\end{align}
where $c \equiv {\lambda h}/{\ell^2}$ denotes the Courant number. Numerical stability of this scheme is guaranteed when the Courant number satisfies $0 \le c \le 1/2$.

\subsection{State Split Fourier Method (SSFM)}

The NLSE involves both linear dispersion terms and nonlinear effects. The state split Fourier method (SSFM) \cite{Agrawal} provides an efficient approach for solving this equation by separating the linear and nonlinear operations. Consider the NLSE:
$
    {\partial A}/{\partial z} = (\hat{D} + \hat{N})A,
$
where $A$ is the complex envelope of the optical field, $\hat{D}$ represents the linear dispersion operator in the frequency domain, and $\hat{N}$ represents the nonlinear operator in the time domain.
The SSFM approximates the solution by alternately applying the dispersion and nonlinear operations over small steps. % $\Delta z$:
%
%\begin{enumerate}
%    \item The dispersion step is computed in the frequency domain:
%	$
%        \tilde{A}(z + \Delta z/2, \omega) = \exp(\hat{D}\Delta z/2)\tilde{A}(z, \omega)
%	$
%
%    \item The nonlinear step is computed in the time domain:
%	$
%        A(z + \Delta z, t) = \exp(\hat{N}\Delta z)A(z + \Delta z/2, t)
%	$
%
%    \item A final dispersion half-step completes the symmetric split:
%    $
%        \tilde{A}(z + \Delta z, \omega) = \exp(\hat{D}\Delta z/2)\tilde{A}(z + \Delta z, \omega)
%	$ 
%\end{enumerate}
%
The method's efficiency comes from utilizing the fast Fourier transform (FFT) to switch between time and frequency domains, where each operator can be applied in its natural domain. The symmetrized version of SSFM achieves second-order accuracy in the step size, making it particularly suitable for our physics-aware decoding framework where accuracy and computational efficiency are both important.
\section{Decoding Problem for PDE Channel}

\subsection{Overview of Channel Model}

Although our proposed method is applicable to various PDEs, we focus on the heat equation (\ref{difusion_PDE}) to illustrate the concept of physics-aware decoding. The boundary conditions for this system are given by:
\begin{align}
u(0,x) &= b(x),\  x \in [0,L], \\
u(t, 0) &= d_1(t),\ 
u(t, L) = d_2(t), \  t \in [0,T],
\end{align}
where $t$ and $x$ represent time and spatial position, respectively. In the following discussion, we present a channel model based on 
this heat conduction process.
Our channel model operates as follows. We begin with a codeword from a binary linear code, $\bm s_0 \equiv (s_1,s_2,\ldots, s_n) \in C(\bm H) \subset \{+1,-1\}^n$, which we represent as a sequence of Gaussian-shaped pulses. This pulse sequence forms the boundary condition $b(x)$ of the heat PDE, effectively embedding our message into the system's initial state at $t=0$. At the receiver, we obtain noisy observations from the PDE solution $u(T,x)$, from which a receiver estimates the transmitted codeword. For brevity, we refer to this system as a PDE channel.

\subsection{Details of Channel Model}

Let $\bm s \equiv (s_1,s_2,\ldots, s_n)^T \in \{+1, -1\}^n$ denote an $n$-dimensional input vector to the PDE channel. To generate the initial waveform for the boundary conditions, we employ a pulse function $\phi: \mathbb{R} \rightarrow \mathbb{R}$. Specifically, we use a Gaussian-shaped pulse function defined as
\begin{align}
\phi(x) \equiv \exp\left(-{x^2}/{(2 T_0^2)} \right),
\end{align}
where $T_0$ represents the half-width of the pulse 
($1/e$-intensity point).
The pulses are centered at positions $p_1,p_2,\ldots, p_n \in [0,L]$. Using these pulse positions, we define the boundary condition function for an input vector $\bm s$ as
\begin{align}
b(x;\bm s) \equiv \sum_{i=1}^n s_i \phi(x - p_i).	
\end{align}
This waveform serves as the transmitted signal and a transmitter embeds the initial waveform $b(x;\bm s)$ into the system as the boundary condition $b(x) = b(x;\bm s)$. We denote the resulting solution of the heat PDE by $u(t,x;\bm s)$. Physically, this transmission process is analogous to selectively heating and cooling positions along an iron wire according to the values in $\bm s$ at time $t = 0$.

At time $t = T$, the receiver employs multiple sensors to measure the values of $u(T,x;\bm s)$. In physical terms, this is equivalent to measuring temperatures at specific positions along the iron wire at time $t = T$. The sensors are located at positions $q_1,q_2,\ldots, q_m \in [0,L]$, and each sensor $i$ provides a received signal $y_i$ given by
\begin{align}
y_i = r_i + n_i, i \in [m],	
\end{align}
where $r_i$ represents the noiseless measurement
\begin{align}
r_i \equiv u(T, q_i;\bm s),\ i \in [m].	
\end{align}
The noise term $n_i$ follows a zero-mean Gaussian distribution with variance $\sigma^2$, i.e., $n_i \sim {\cal N}(0, \sigma^2)$. Th receiver effectively samples the PDE solution $u(t,x;\bm s)$ at the sensor positions, subject to additive white Gaussian noise.

\begin{figure}[htbp]
\begin{center}
\includegraphics[width=0.9\columnwidth]{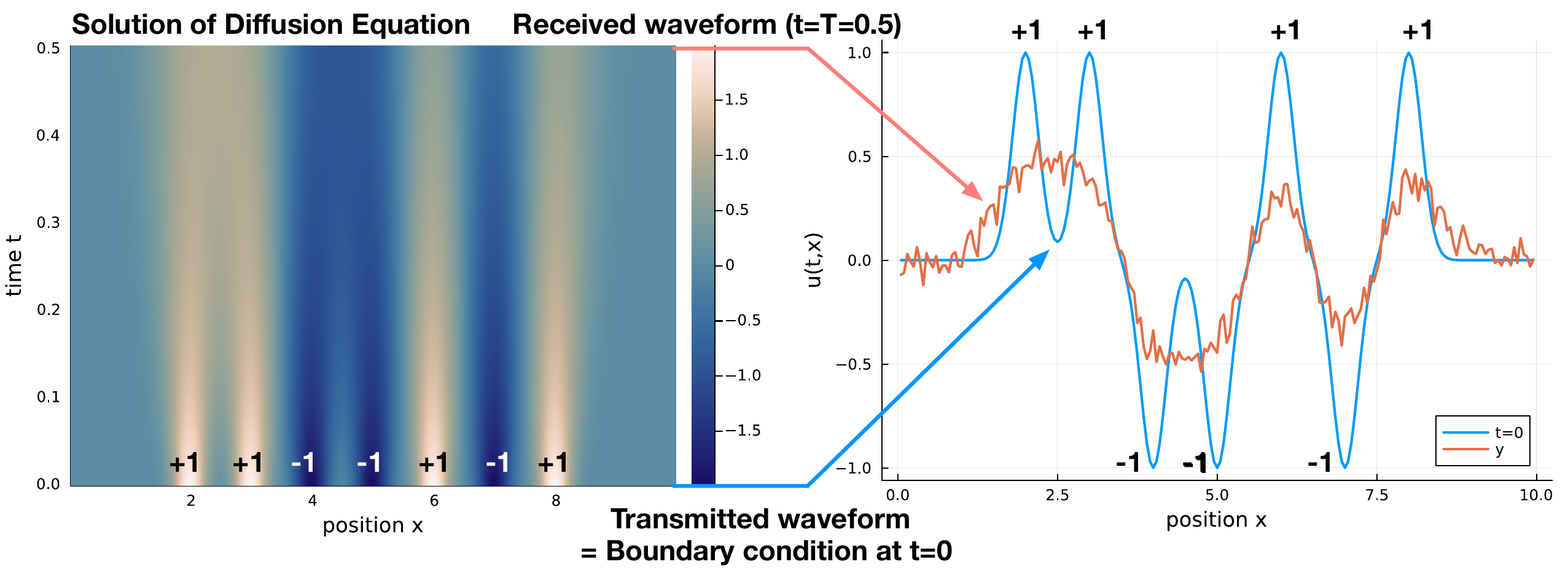}
\caption{A PDE channel defined by a heat PDE with $\lambda = 0.2$. A bipolar vector $\bm s = (+1,+1,-1,-1,+1,-1,+1)$ is used as the input vector and Gaussian-shaped pulses are used for generating input waveform.}
\label{fig:channel_model}
\end{center}
\end{figure}

Figure \ref{fig:channel_model} demonstrates the behavior of our PDE channel. The left panel shows the evolution of the solution $u(t,x;\bm s)$ for the heat PDE with parameters $\lambda = 0.2$, $T = 0.5$, and $L = 10$. We used a seven-bit bipolar vector $\bm s = (+1,+1,-1,-1,+1,-1,+1) \in \{+1,-1\}^7$ as the input and solved the PDE numerically using FDM with a $200 \times 100$ grid ($x$ direction $\times$ $t$ direction). The solution reveals how the initial waveform at the bottom gradually diffuses over time, resulting in progressive signal blurring as $t$ increases.
The right panel compares the initial waveform at $t = 0$, generated from the input vector $\bm s$, with the received waveform at $t = T$ under noise conditions of $\sigma = 0.05$. The received signal exhibits  a low-pass filtered version of the input signal with additive Gaussian noise, which is consistent with the conduction process acting as a low-pass filter.

\subsection{Decoding Problem}

Consider the heat PDE, which admits a unique solution. The relationship between the noiseless received vector $\bm r \equiv (r_1,\ldots, r_m)$ and the input vector $\bm s$ can be characterized by a deterministic function $f: \mathbb{R}^n \rightarrow \mathbb{R}^m$, such that $\bm r = f(\bm s)$. This allows us to express our observation model compactly as
\begin{align}
\bm y = f(\bm s) + \bm n,	
\end{align}		
where $\bm y \equiv (y_1, y_2,\ldots, y_m)^T$ represents the received signal and $\bm n \equiv (n_1, n_2,\ldots, n_m)^T$ denotes the additive white Gaussian noise vector.
Our objective is to estimate the input vector $\bm s$ with maximum accuracy. Given that $\bm s$ is uniformly distributed over a bipolar code $C(\bm H)$ and the noise follows a Gaussian distribution, the maximum likelihood (ML) decoding problem is formulated as
\begin{align}
\hat{\bm s} = \mbox{arg} \min_{\bm s \in {C(\bm H)}} \| \bm y - f(\bm s) \|^2.	
\end{align}
However, this ML decoding is computationally prohibitive, as it requires an exhaustive search over the codebook $C(\bm H)$ whose size grows exponentially with $n$.

\section{Proposed Method}

\subsection{Overview of Physics-Aware Decoding}

The start point of our proposal is to use discretized version of GF decoding \cite{wadayama24-2}:
\begin{align} \label{GFGD}
	\bm s^{(k+1)} &= \bm s^{(k)} - \eta (\nabla_{\bm s^{(k)}} L(\bm s^{(k)};\bm y ) + \gamma \nabla_{\bm s^{(k)}} h_{\alpha,\beta}(\bm s^{(k)}))
\end{align}
for approximating the ML decoding above where 
$\eta$ is a step size parameter. We can replace the gradient of the negative log likelihood 
$\nabla L (\bm x; \bm y)$ by $\nabla_{\bm x} \| \bm y - f(\bm x) \|^2$ 
because 
\begin{align}
\nabla L (\bm x; \bm y) \propto \nabla_{\bm x} \| \bm y - f(\bm x) \|^2.	
\end{align}
Of course, no concise representation of $\nabla_{\bm x} \| \bm y - f(\bm x) \|^2$ is available because
the function $f$ involves a state evolution process defined by a heat PDE.
We thus approximate $f$ by an approximate function $\tilde f$ obtained by 
a differentiable PDE solver.
Namely, let 
\begin{align}
	\tilde r_i \equiv \tilde u(T, q_i;\bm s),\ i \in [m]	
\end{align}
and 
\begin{align}
	\tilde{\bm r} = \tilde{f}(\bm s),	
\end{align}
where $\tilde{\bm r} = (\tilde{r}_1,\tilde{r}_2,\ldots, \tilde{r}_m)^T$.
The approximate solution $\tilde u(T, x;\bm s)$ is given from the PDF solver.
By replacing $f$ by $\tilde f$, we have a computationally tractable recursive formula:
\begin{align} \label{GFGD3}
	\bm s^{(k+1)} &= \bm s^{(k)} - \eta (\nabla_{\bm s^{(k)}} \| \bm y - \tilde f(\bm s^{(k)}) \|^2 + \gamma \nabla_{\bm s^{(k)}} h_{\alpha,\beta}(\bm s^{(k)})).
\end{align}
It is important to note that the advantage of this formulation is that we can utilize AD to compute the gradient $\nabla_{\bm s} \| \bm y - \tilde f(\bm s) \|^2$. 
\begin{figure}[htbp]
\begin{center}
\includegraphics[width=0.9\columnwidth]{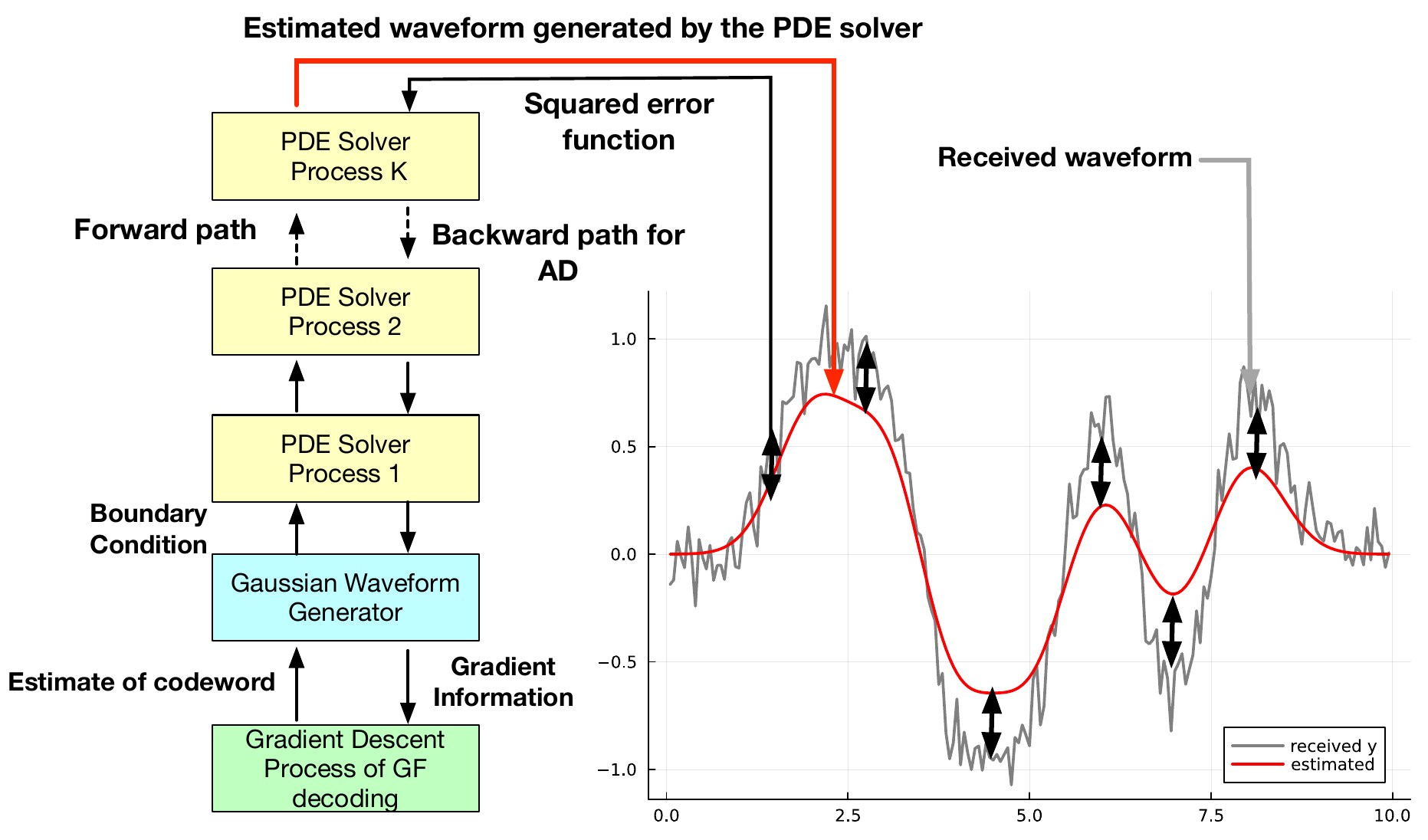}
\caption{Block diagram of physics-aware decoding.}
\label{fig:decoding}
\end{center}
\end{figure}

Figure \ref{fig:decoding} illustrates the block diagram of physics-aware decoding. 
A GF decoder executes a gradient descent process 
according to
(\ref{GFGD3}) and generates an estimated codeword $\hat{\bm s}$,
which is passed to Gaussian waveform generator to set up the boundary condition. 
Then, a PDE solver produces an approximate solution $\tilde f({\bm s})$.
	The squared error $\|\bm y - \tilde f({\bm s})\|^2$ is
	used as a loss function and the gradient information 
	is fed back to the PDE solver. The AD mechanism helps 
	backward signal propagation and a GF decoder finally obtain
		the gradient information $\nabla_{\bm s} \|\bm y - \tilde f({\bm s})\|^2$. This gradient information is 
		then used for updating the codeword estimate.

\subsection{Details of Physics-Aware Decoding}

Details of the {physics-aware decoding}
are presented in Algorithm \ref{GF_alg} for the 
heat PDE. We assume the grid of size 
$N_x \times N_t$. Each grid cell 
has the size $\ell \times h$.
\begin{algorithm}[htbp]
 \caption{Physics-Aware Decoding}
 \label{GF_alg}
 \begin{algorithmic}[1]
  \STATE Sample the initial state $\bm s^{(0)}\sim {\cal N}(0, \sigma_s^2)$.
  \FOR {$k := 0$ to $U-1$}
\STATE Set the initial vector $\bm u{[0]} \equiv (u{[0]}_1,\ldots, u{[0]}_{N_x-1})$ for the PDE solver by 
\begin{align}
	u{[0]}_i := b(i\ell;\bm s^{(k)}),\quad i \in [N_x - 1].
\end{align}
\STATE Solve the PDE with a PDE solver and 
generate $\tilde r_i := \tilde u(T, q_i;\bm s^{(k)}),\ i \in [m]$.
%  \FOR {$\tau := 0$ to $N_t-1$}
%\STATE Execute a FDM process:
%\begin{align}	
%\bm u^{(\tau+1)} := \bm P \bm u^{(\tau)} + c\bm d^{(\tau)}.
%\end{align}
%\ENDFOR
%\STATE Set $\tilde{\bm r} \equiv (\tilde r_1,\tilde r_2,\ldots, \tilde r_m)$ by
%$
%	\tilde r_i := u^{(N_t)}_{ \lfloor q_i/l \rfloor },\ i \in [m].
%$
\STATE Compute the gradient of the loss function by AD:
\begin{align}
\bm z \equiv (z_1,\ldots, z_{Nx-1})^T&:=\nabla_{\bm u^{[0]}} \| \bm y - \tilde{\bm r} \|^2.
\end{align}
\STATE Let  $\bm g := (g_1,g_2,\ldots, g_n)^T$ be
$
	g_i := z_{\lfloor p_i/\ell\rfloor},\quad i \in [n].
$
\STATE Compute the gradient of the code potential energy:
\begin{align} 
\bm d_{abs} &:= \exp(\boldsymbol H\ln(|\bm s^{(k)}|))\\
\bm d_{sgn} &:= \boldsymbol 1-2\mathrm{bmod}(\boldsymbol {H}(\boldsymbol 1-\mathrm{sgn}(\bm s^{(k)})/2)) \\
\boldsymbol d &:= \bm d_{sgn}\odot \bm d_{abs}\\ \nonumber
\bm h&:=4 \alpha (\bm s^{(k)}\odot\bm s^{(k)}-\boldsymbol 1)\odot\bm s^{(k)} \\
	&+2\beta\boldsymbol{H}^{T}(\boldsymbol d\odot\boldsymbol d-\boldsymbol d)/\bm s^{(k)}. 
\end{align}
\STATE 	Execute the gradient descent process of GF decoding:
\begin{align}
\bm s^{(k+1)} := \bm s^{(k)} - \eta (\bm g + \gamma \bm h).
\end{align}
  \ENDFOR
\STATE Output $\hat{\bm s} := \mbox{sgn}(\bm{s}^{(U)})$.
 \end{algorithmic} 
 \end{algorithm}
It should be remarked that the gradient 
$
\nabla_{\bm s} \| \bm y - \tilde f(\bm s) \|^2
$
that is required for the gradient descent process 
is approximated by the values of 
$\nabla_{\bm u^{(0)}} \| \bm y - \tilde f(\bm s) \|^2$ at $p_i (i \in [n])$. 
This approximation greatly reduces the computational
complexity for backward gradient computation because we can skip 
the backward path of the Gaussian waveform generator of Algorithm \ref{GF_alg}. From the results of numerical experiments, even with 
this approximation, Algorithm \ref{GF_alg} works well as expected.

%In this context, we assumed the heat PDE (\ref{difusion_PDE}) and 
%the corresponding FDM solver described in Subsection \ref{sec:FDM_solver}.
%In principle, other type of PDE and a differentiable numerical solver 
%can be used for physics-aware decoding. What we need is to replace 
%Step 5 of Algorithm \ref{GF_alg} with a suitable solver.
We employed the squared error function as the measure the discrepancy between 
$\bm y$ and $\tilde{\bm r}$ because additive white Gaussian noises are assumed in the channel model.
If the noises are not Gaussian, we may be able to replace the error function 
according to the noise statistics.

%The proposed algorithm is a double-loop algorithm consisting of 
%the outer gradient descent loop and the inner PDE solver loop.
%In a gradient loop, the dominant computation is the PDE solver loop 
%that needs $O(N_t N_x)$ computation for an iterative solver.
%If we use an LDPC code, the computational complexity of the gradient of the code potential energy 
%is bounded by $O(n)$ where $n < N_x$ in general.
%Since the gradient computation by
%autograd needs also $O(N_t N_x)$-time, 
%the time complexity 
%of the whole algorithm becomes $O(U N_t N_x)$.
%In order to reduce the computation time, 
%one can use coarse grid, i.e., smaller $N_t$ and $N_x$,
%but it may cause the degradation of the solution quality. The choice of the grid size may be crucial to balance the computation complexity and the decoding 
%performance.

\section{Numerical Experiments }

\subsection{Heat Equation}

\begin{figure}[htbp]
\begin{center}
\includegraphics[width=0.9\columnwidth]{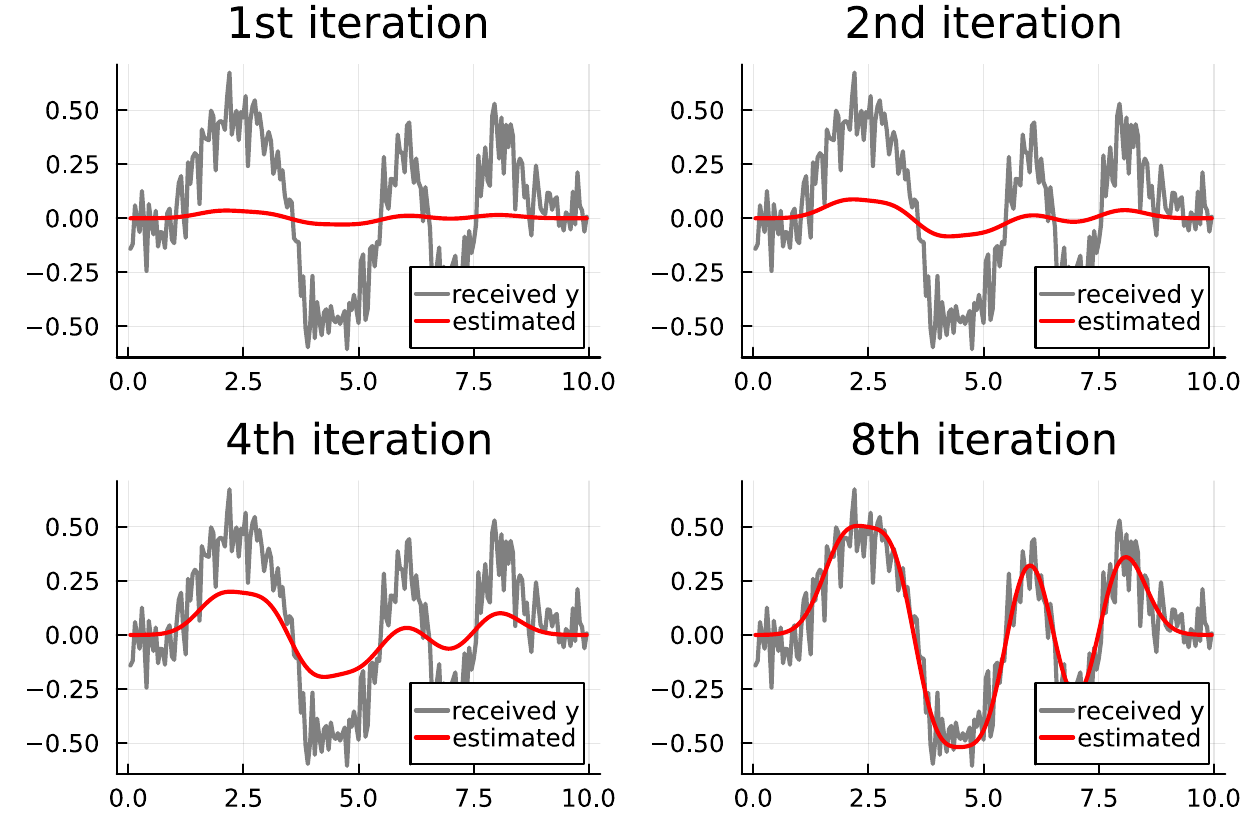}
\caption{Received waveform and estimated output waveform $\bm u^{(N_t)}$ in a decoding process.}
\label{fig:evolving_output}
\end{center}
\end{figure}

In this subsection, a decoding process of a numerical experiment is
demonstrated for the heat equation.
Figure \ref{fig:evolving_output} 
presents evolutions of estimated output waveform ${\bm{u}}{[N_t]}$ 
in a decoding process.
%In the case of diffusion PDE, we can greatly simplifies the PDF solver.
%Let $(N_x-1)\times (N_x-1)$ real matrix $\bm P$ be 
%\begin{align}
%\bm P \equiv
%\left(
%\begin{array}{cccccc}
%1-2c & c & 0 &  \cdots  & 0& 0	\\
%c & 1-2c & c &   \cdots  & 0 & 0 \\
%0 & c    & 1-2c & \cdots  & 0& 0\\
%\vdots & \vdots &  \vdots & \vdots & \vdots &\vdots\\
%0 & 0 & 0 &  \cdots  & 1-2c & c \\
%0 & 0 & 0 &  \cdots  & c & 1-2c \\
%\end{array}
%\right),
%\end{align}
%where $N_x$ represents the number of grids in position direction.
%When the the boundary condition function $d_1(t) = d_2(t) = 0$ for $t \in [0,T]$.
%the FDM solver, in this case can become a simple matrix-vector product:
%$\bm u^{(N_t)} = \bm Q \bm u^{(0)}$ where $Q \equiv P^{N_t}$ and  $N_t$ 
%represents the number of grids in time direction. In this case, the gradient 
%is efficiently evaluated by 
%$\bm z = \nabla_{\bm u^{(0)}} \| \bm y - \tilde{\bm r} \|^2 = \bm Q^T(\bm Q \bm u^{(0)}- \bm y)$.
%In the following experiment, we used this matrix-based FDM.
The heat PDE is solved by the FDM.
The following parameters were used.
We used $200 \times 100$ ($x$ direction $\times$ $t$ direction) grid
with grid size $h = 0.005$ and $\ell = 0.05$.
The heat PDE with $\lambda = 0.2$ was solved in the 
region $[0,T =0.5] \times [0,L = 10]$.
The noise standard deviation was set to $\sigma = 0.1$.
The sensor positions were all the grid points $q_i = \ell i (i \in [N_x - 1])$.
A codeword of (7,4) Hamming code was chosen as a transmitted word.
In all the experiments in this subsection,
to generate received signals, 
we synthesized the received waveform using the FDM solver.
The step size of the gradient descent was set to $\eta = 0.1$ and
the parameter $\gamma$ was set to $1.0$.
From Fig.~\ref{fig:evolving_output}, we can see that the estimated 
waveform gradually approaches to the received signal $\bm y$.
This means that the gradient descent process works appropriately to 
reduce the squared error between $\bm y$ and $\bm u^{(N_t)}$.
%From Fig.~\ref{fig:evolving_input}, we can also confirm that 
%the estimate of $\bm s$ becomes accurate as iteration proceeds.
%\begin{figure}[htbp]
%\begin{center}
%\includegraphics[width=\columnwidth]{../experiments/exp-2/evolving_input.pdf}
%\caption{Input waveform and estimated input waveform $\bm u^{(0)}$.}
%\label{fig:evolving_input}
%\end{center}
%\end{figure}

\begin{figure}[htbp]
\begin{center}
\includegraphics[width=0.9\columnwidth]{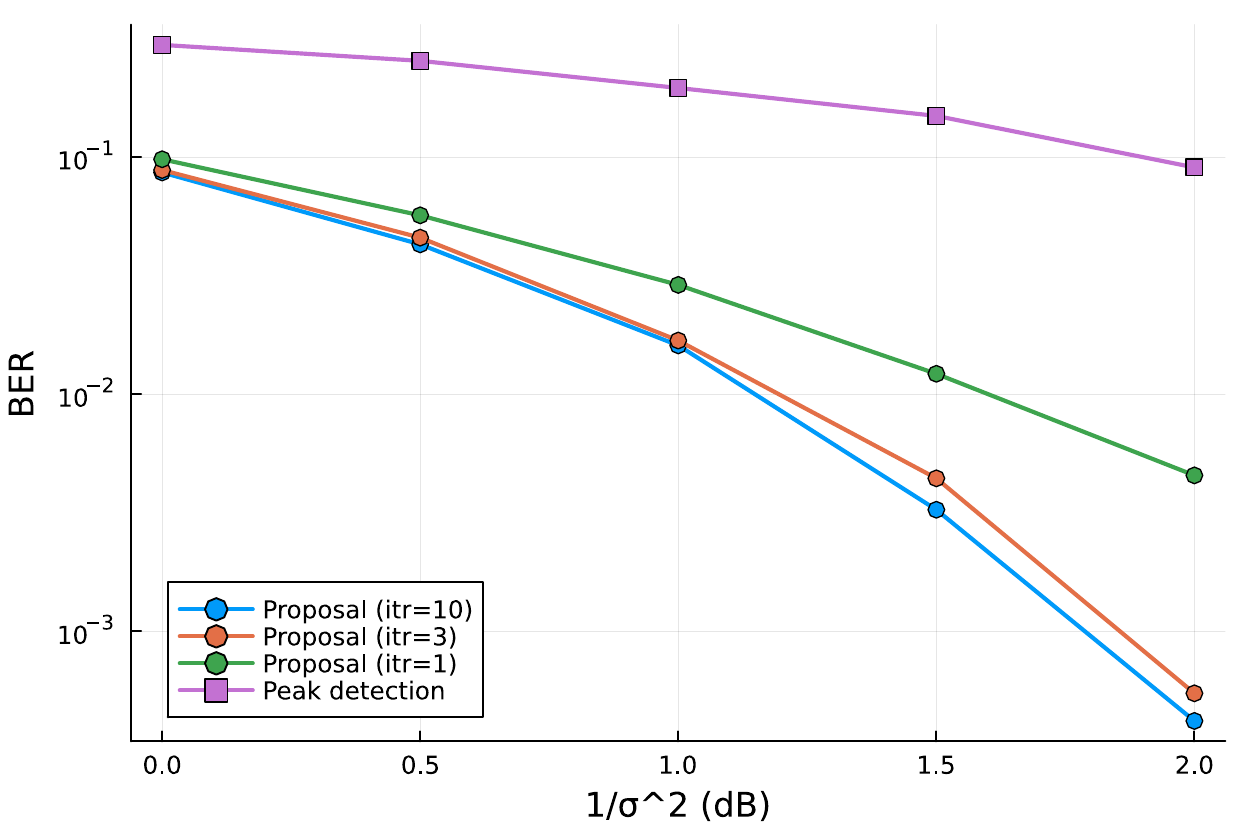}
\caption{BER performance of the proposed algorithm for the PDE channel governed by the heat PDE. The (31,15) BCH code is used. The error curve of peak detection method is also included as the baseline. %(Right) A sample of an input waveform and received waveform $1/\sigma^2 = 2$(dB). 
}
\label{fig:BER_curve}
\end{center}
\end{figure}

Figure \ref{fig:BER_curve} show the bit error rate (BER) obtained by a numerical experiment.
The parameters setting for the FDM solver were as follows:
$h = 0.001, l = 0.01, \lambda = 0.01, N_x = 512, N_t = 50, T=0.05, L = 5.12$ and $T_0 = 0.02$. The parameters for GF decoding were: $\eta = 0.1, \gamma = 0.1, \alpha = \beta = 1$ and the number of iterations was set to 20.
As a baseline scheme, we used 
the {peak detection method} which simply uses the polarity of $\bm y$ at the sensor positions to produce a bipolar estimate. 
We used the bipolar version of the $(31,15)$ BCH code
in this experiment. From Fig.~\ref{fig:BER_curve}, we can observe that the proposed method achieves much steeper decrement compared with the baseline curve.
This indicates that the proposed method can effectively utilize the redundancy provided by the BCH code.

\subsection{Nonlinear Schrödinger Equation (NLSE)}
In this subsection, we study the normalized NLSE
\begin{align} \label{NLE}
	\frac{\partial U}{\partial \xi}  = - \frac{i s}{2} \frac{\partial^2 U}{\partial \tau^2} + i N^2 |U|^2 U  
\end{align}
in the context of optical fiber communications \cite{Agrawal},
where $i$ represents the unit of imaginary number.
The variables $\xi$ and $\tau$ represent 
the normalized position and time, respectively. The function $U$ is the normalized optical field. This PDE describes how an input waveform evolves in a single mode optical fiber. 
The parameter $s$ is given by $s \equiv \mbox{sgn}(\beta_2)$ 
where $\beta_2$ is the dispersion constant.
The parameter $N^2$ relates the nonlinearity of the optical fiber. The PDE (\ref{NLE}) contains the nonlinear term  $i N^2 |U|^2 U$, which causes nonlinear distortion.

In the following numerical experiment, the parameters were set to 
$s = 1$ and $N^2 = 1$.  The linear dispersion length 
$L_D$ was set to $0.1$.
The number of grids for time direction was set to 256. 
The observation of $\bm y$ is conducted at $\xi = 0.5$. 
The grid length for $\xi$-direction was set to $\ell = 0.025$.
As a PDE solver, we used the SSFM solver 
to generate the received signals.

The simulation code is implemented using 
Julia 1.9 and AD in Zygote.jl.
The parameter setting of GF decoder is as follows.
The parameters $\eta$ and $\gamma$ were set to $0.1$,
and $\alpha$ and $\beta$ were set to 1.
The number of iteration was fixed to 10 and 20.
As a baseline, we employed 
{\em Backpropagation method (BP)} \cite{Ip}
which is a well-known nonlinearity compensation technique 
in optical fiber communications. BP processes the received signal 
by numerically solving the NLSE in the reverse direction 
with opposite signs of dispersion and nonlinearity coefficients. 
%This can be viewed as ``undoing'' the channel effects by propagating the signal backward through a virtual fiber with inverse channel parameters. 

\begin{figure}[htbp]
\begin{center}
\includegraphics[width=0.9\columnwidth]{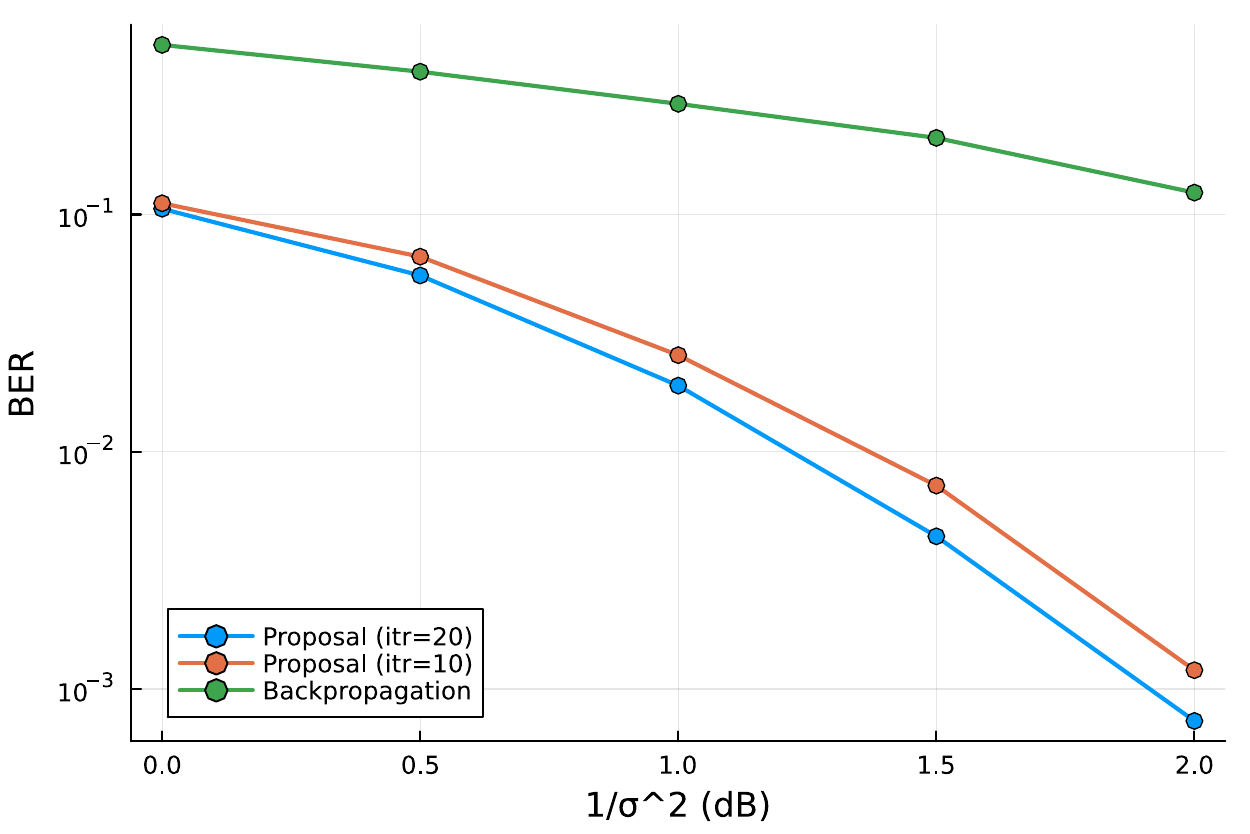}
\caption{BER performance of the proposed algorithm for the PDE channel governed by 
the NLSE (\ref{NLE}). (15,7) BCH code is used. Conventional BP method is used as a baseline. 
}
\label{fig:BER_NLS}
\end{center}
\end{figure}

Figure \ref{fig:BER_NLS} shows the BER of the proposed algorithm. The BER 
performance of the proposed algorithm with the number of iteration 20 
significantly outperforms that of the
baseline.
It is worth noting that BP can theoretically achieve perfect signal reconstruction in the absence of noise, as it exactly inverts the channel effects by reverse-propagating the signal. However, in the presence of noise, BP suffers from significant noise enhancement, similar to the noise amplification observed in zero-forcing equalization for linear channels. 
The proposed algorithm can avoid such noise enhancement that is a clear advantage over BP-based nonlinear compensation techniques \cite{Ip}.

\section{Concluding Summary}

This paper has introduced a novel physics-aware decoding framework that integrates PDE-based channel modeling with error correction techniques. By combining differentiable PDE solvers with GF decoding, we have demonstrated improved decoding performance through numerical experiments on both the heat equation and the NLSE. The key innovation lies in the seamless integration of physical models with error correction techniques through AD, enabling the decoder to utilize gradient information from the PDE solver effectively.

The primary future challenge lies in the computational complexity introduced by the double-loop structure of our method. Although using coarser grids can partially address this issue, balancing computational efficiency with solution accuracy remains a crucial challenge for practical implementations. Nevertheless, our findings suggest that physics-aware decoding opens promising new avenues in communication system design, with applications potentially extending beyond decoding to general signal detection and signal recovery tasks.


\begin{thebibliography}{00}

	\bibitem{Agrawal}
	G.~P.~Agrawal,
	``Nonlinear fiber optics (6th ed.),'' Academic Press, 2019.

	\bibitem{Farsad}
	N. Farsad, H. B. Yilmaz, A. Eckford, C. -B. Chae and W. Guo, ``A comprehensive survey of recent advancements in molecular communication,'' in IEEE Communications Surveys \& Tutorials, vol. 18, no. 3, pp. 1887-1919, 2016.
	
	\bibitem{Gallager63}
  R. G. Gallager,
  ``Low density parity check codes,''
	MIT Press, 1963.

	\bibitem{Raissi}
	M. Raissi, P. Perdikaris and G.E. Karniadakis
	``Physics-informed neural networks: a deep learning framework for solving forward and inverse problems involving nonlinear partial differential equations,'' Journal of Computational Physics, Vol. 378, pp.686-707, 2019.
	
	\bibitem{Baydin2018}
A.~G. Baydin, B.~A. Pearlmutter, A.~A. Radul,  and J.~M. Siskind, ``Automatic
  differentiation in machine learning: a survey,'' \emph{Journal of Machine
  Learning Research}, vol.~18, pp. 5595--5637, 2018.

	\bibitem{Bartels}
	S.~Bartels,
	``Numerical Approximation of Partial Differential Equations,'' Springer International Publishing Switzerland, 2016.

    \bibitem{wadayama23}T. Wadayama, K. Nakajima, and A. Nakai-Kasai, ``Gradient flow decoding for LDPC codes,'' 2023 International Symposium on Topics in Coding (ISTC2023), Brest, France, 2023. 

    \bibitem{wadayama24-2}T. Wadayama and L. Wei, ``Generalized gradient flow decoding and and Its Tensor-Computability,'' International Symposium on Information Theory (ISIT2024), Athens, 2024. 

    \bibitem{wadayama10}T. Wadayama, K. Nakamura, M. Yagita, Y. Funahashi, S. Usami, I. Takumi, ``Gradient descent bit flipping algorithms for decoding LDPC codes'', IEEE Trans. Comm., pp.1610-1614, vol.58, no.6, June (2010)

%	\bibitem{wadayama23b}
%    T. Wadayama and S. Takabe, ``Proximal decoding for LDPC codes,'' IEICE Transactions on Fundamentals of Electronics, Communications and Computer Sciences, vol. E106-A, no. 3 pp. 359-367 (2023). 

   	\bibitem{Feldman03}
      J. Feldman, ``Decoding error-correcting codes via linear programming,''
 		Massachusetts Institute of Technology, Ph. D. thesis, 2003.

 	 \bibitem{Vontobel08}
     P. O. Vontobel,
     ``Interior-point algorithms for linear-programming decoding,''
     IEEE Information Theory and Applications Workshop, 2008.
     
         \bibitem{Zhang13}
  	X. Zhang and P. H. Siegel, 
    ``Efficient iterative {LP} decoding of {LDPC} Codes with alternating direction method of multipliers,''
    IEEE International Symposium on Information Theory (ISIT), 2013.

	\bibitem{Barman13}
  S. Barman, X. Liu, and S. C. Draper, and B. Recht,
  ``Decomposition methods for large scale LP decoding,''
  IEEE Transactions on Information Theory,
  vol.59, no.12, pp.7870-7886, 2013.

	\bibitem{Ip}
	E. Ip and J. M. Kahn, ``Compensation of dispersion and nonlinear impairments using digital backpropagation,'' in Journal of Lightwave Technology, vol. 26, no. 20, pp. 3416-3425, 2008.
     
     \bibitem{LISTA}
K. Gregor, and Y. LeCun, ``Learning fast approximations of sparse coding,'' Proc. 27th Int. Conf. Machine Learning, pp. 399-406, 2010.

\bibitem{9020494}
A.~Balatsoukas-Stimming and C.~Studer, ``{Deep unfolding for communications
  systems: a survey and some new directions},'' in \emph{\textit{Proc.} 
  IEEE International Workshop on Signal Processing Systems (SiPS)}, pp.
  266--271, 2019.
  
\bibitem{wadayama-2025-1}
T.~Wadayama, K.~Igarashi, and T.~Takahashi,
``Physics-aware sparse signal recovery through PDE-governed measurement systems, ''
submitted to ISIT 2025.


\end{thebibliography}
\end{document}